\documentclass[useAMS,usenatbib]{mn2e}

\usepackage{hyperref}

\providecommand{\adsurl}[1]{\href{#1}{ADS}}
 
\providecommand{\url}[1]{\href{#1}{#1}}

\newcommand\eq[1]{Eq.~(\ref{#1})}
\newcommand\eqs[2]{Eqs.~(\ref{#1}) and (\ref{#2})}

\newcommand{\beq}{\begin{equation}}
\newcommand{\eeq}{\end{equation}}
\renewcommand{\d}{{\rm d}}
\renewcommand{\P}{{\rm Pr}}
\newcommand{\tobs}{t_{\rm obs}}
\newcommand{\p}{\wp}
\newcommand{\e}{{\rm e}}
\newcommand{\bB}{{\bar{B}}}
\newcommand{\N}{\mathcal{N}}
\newcommand{\n}{{\bf \hat{n}}}
\newcommand{\D}{{\bf \hat{d}}}
\newcommand{\dchieff}{\Delta\chi^{2}_{\rm eff}}
\newcommand{\Lmax}{\mathcal{L}_{{\rm max}}}

\newcommand{\lsim}{\,\raise 0.4ex\hbox{$<$}\kern -0.8em\lower 0.62ex\hbox{$\sim$}\,}

\begin{document}

\title[Bayesian Calibrated Significance Levels]{
Bayesian Calibrated Significance Levels Applied to the Spectral Tilt and Hemispherical Asymmetry }

\author[Gordon and Trotta]{Christopher Gordon and Roberto Trotta\\
Oxford University, Astrophysics Department, Denys Wilkinson
Building, Keble Road, Oxford, OX1 3RH, UK }

\date{}
\maketitle
\begin{abstract}
Bayesian model selection provides a formal method of determining
the level of  support for new parameters in a model. However, if
there is not a specific enough underlying physical motivation for
the new parameters it can be hard to assign them meaningful
priors, an essential ingredient of Bayesian model selection. Here
we look at methods maximizing the prior so as to work out what is
the maximum support the data could give for the new parameters. If
the maximum support is not high enough then one can confidently
conclude that the new parameters are unnecessary without needing
to worry that some other prior may make them significant. We
discuss a computationally efficient means of doing this which
involves mapping p--values onto upper bounds of the Bayes factor
(or odds) for the new parameters. A p--value of 0.05
($1.96\sigma$) corresponds to odds less than or equal to 5:2 which
is below the `weak' support at best  threshold. A p--value of
0.0003 ($3.6\sigma$) corresponds to odds of less than or equal to
150:1 which is the `strong' support at best threshold. Applying
this method we find that the  odds on the scalar spectral index
being different from one are 49:1 at best. We also find that the
odds that there is primordial hemispherical asymmetry in the
cosmic microwave background are 9:1 at best.
\end{abstract}

\section{Introduction}

When there are several competing theoretical models, Bayesian
model selection provides a formal way of evaluating their relative
probabilities in light of the data and any prior information
available. A common scenario is where a model is being extended by
adding new parameters. Then the relative probability of the model
with the extra parameters can be compared with that for the
original model. This provides a way of evaluating whether the new
parameters are supported by the data. Often the original model is
``nested'' in the new model in that the new model reduces to the
original model for specific values of the new parameters.
The Bayesian framework automatically implements an
Occam's razor effect as a penalization factor for less
predictive models  -- the best
model is then the one that strikes the best balance between
goodness of fit and economy of parameters~\citep{trotta05}.

For nested models, the Occam's razor effect is controlled by the
volume of parameter space enclosed by the prior probability
distributions for the new parameters.
The  relative probability of the new model can be made arbitrarily
small by increasing the broadness of the prior. Often this is not
problematical as prior ranges for the new parameters can (and
should) be motivated from the underlying theory. For example, in
estimating whether the scalar spectral index ($n$) of the
primordial perturbations is equal to one (see Sec.~\ref{sec:n}),
the prior range of the index can be constrained to be $0.8\lsim n
\lsim 1.2$ by assuming the perturbations were generated by slow
roll inflation. The sensitivity of the model selection result can
also be easily investigated for other plausible, physically
motivated choice of prior ranges (e.g.,
\cite{Trotta:2007hy,Trotta:2006ww}).

However, there are cases like the asymmetry seen in the WMAP
cosmic microwave background (CMB) temperature data (see
Sec.~\ref{sec:as}) where there is not a specific enough model
available to place meaningful limits on the prior ranges of the
new parameters. This hurdle arises frequently in cases when the
new parameters are a phenomenological description of a new effect,
only loosely tied to the underlying physics, such as for example
expansion coefficients of some series. In these cases, an
alternative is to choose the prior on the new parameters in such a
way as to maximise the probability of the new model, given the
data. If, even under this best case scenario, the new model is not
significantly more probable than the old model, then one can
confidently say that the data does not support the addition of the
new parameters, regardless of the prior choice for the new
parameters.



\section{Upper Bounds on the Bayes factor}
\label{sec:bounds}

A model ($M_0$) may be compared to a new model with extra
parameters ($M_1$)  using the Bayes factor (also known as the odds)
 \beq B = { p(x|M_1)
 \over p(x|M_0)}, \eeq
where $x$ is the data and the model likelihood $p(x|M_i)$
$(i=0,1)$ is given by
 \beq
 p(x|M_i) = \int \d \theta_i p(x|\theta_i, M_i) p(\theta_i|M_i)
 \eeq
with $\theta_i$ denoting the parameters under model $M_i$. The
Bayes factor gives the change in the relative probability of the
two models brought about by the data $x$, i.e.
 \beq
 \frac{p(M_1|x)}{p(M_0|x)} = B \frac{P(M_1)}{P(M_0)},
 \eeq
where $P(M_i)$ ($i=0,1$) are the prior probabilities for the two
models and $P(M_i|x)$ the posterior probabilities. The level of
support is usually categorized as either `inconclusive' ($|\ln
B|<1$), `weak' ($1\le |\ln B| \le 2.5$), `moderate' ($2.5\le|\ln
B|\le 5$), or `strong' ($|\ln B| \ge 5$).

We denote the new parameters by $\theta$ and they are fixed to be
$\theta^*$
under the simpler model (we restrict our considerations to nested
models). The Bayes factor is then, using a generalized version of
the Savage--Dickey density ratio (see \cite{trotta05} for details)
 \beq B = \frac{p(\theta^*|M_1)}{p(\theta^*|x, M_1)},
\label{B}
 \eeq
 where $p(\theta^*|x, M_1)$ is the posterior
distribution under $M_1$, evaluated at $\theta=\theta^*$. If
$p(\theta|M_{1})$ is made sufficiently broad, $p(\theta^*|x, M_1)$
depends only on the likelihood. Thus, $B$  can be made arbitrarily
small by making $p(\theta|M_1)$ sufficiently broad (since the
prior must be normalized to unity probability content, a broader
$p(\theta|M_1)$ corresponds to a smaller value of
$p(\theta^*|M_1)$). This is not problematical if the physical
model underlying $M_1$ is specific enough to provide some
well--motivated prior bounds on $\theta$. When this is not the
case, an upper bound can still be obtained on $B$ by optimizing
over all priors and choosing $p(\theta|M_{1})$ to be a delta
function centered at the maximum likelihood value, $\theta_{\rm
max}$. This is the choice that maximally favours $M_1$, and the
upper bound on the odds is then
 \beq
\bar{B}={p(x|\theta_{\rm max}, M_1) \over p(x|\theta^{*}, M_0)},
 \eeq
corresponding to the likelihood ratio between $\theta_{\rm max}$
and $\theta^*$. However,
such a choice of prior fails to capture that $M_{1}$ is supposed to be a more complex model than $M_{0}$. Since $\theta^{*}$ represents the 
theoretically motivated simpler hypothesis, it makes sense that the alternative hypothesis has a more spread out prior distribution for $\theta$.
Furthermore, if there is {\em a priori} no
strong preference for either $\theta>\theta^{*}$ or
$\theta<\theta^{*}$, then it may be preferable to maximize over
priors that are symmetric about $\theta^{*}$ and unimodal
 (the
latter requirement coming again from a principle of indifference).
 \citet{bersel87} show that maximizing $B$ over all such $p(\theta
|M_1)$ is the same as maximizing over all $p(\theta |M_1)$ that
are uniform and symmetric about $\theta^*$. We give an explicit example of this procedure in \eq{example}.

However, this optimization may be computationally prohibitive as
 evaluating \eq{B} usually  requires numerical evaluation of high dimensional integrals
\citep{mukparlid06,Feroz:2007kg}. An alternative way of obtaining
an upper bound on $B$, that does not rely on explicitly specifying  a class of alternative priors for $\theta$, is to use Bayesian calibrated p--values
\citep{selbayber01}. First, frequentist methods are used to obtain
the p--value. To do this, a test statistic $(t)$ needs to be
chosen, with the general property that the larger the value the
less well the data agree with $M_0$. A common choice is the
improvement in the maximum likelihood value when the additional
parameters are allowed to vary. However, if the likelihood is
computationally expensive to obtain, then other measures may be
used. The p--value is given by
 \beq
 \p=p(t\ge \tobs(x) | M_0),
  \eeq
  where $\tobs(x)$ is the value of
$t$ estimated from the data. The key property of p--values is that
if $M_0$ is correct, and $t$ is a continuos statistic, then the
probability distribution of $\p$ will be uniform, $p(\p|M_0)=1$
for $0 \le \p \le 1$. The final result will not be sensitive to
the precise choice of $t$. The only property needed for $t$ is
that it should be a continuous statistic and larger values of $t$
should correspond to less agreement with $M_0$. It follows that
$p(\p|M_1)$ will be monotonically decreasing for $ 0 \le \p \le
1$. \citet{selbayber01} express the Bayes factor in terms of the
distribution of the p--values \beq B={p(\p | M_{1}) \over
p(\p|M_{0}) }=  p(\p|M_{1}) \, . \eeq They look at a wide range of
non-parameteric monotonically decreasing distributions for
$p(\p|M_1)$ and under mild regularity conditions, they find the
upper bound
 \beq B\le \bB=
{ -1 \over \e\p\ln \p}
\label{calib}
 \eeq for $\p\le\e^{-1}$, where $\e$ is the
exponential of one. Table~\ref{translation} lists $\bB$
for some common thresholds of $\p$ and $\ln B$.
\begin{table}
\begin{center}
\begin{tabular}{l|r|r|r|l }
p--value                    &  $\bB$                 &  $\ln \bB$
&sigma&category\\ \hline
 0.05 & 2.5 & 0.9 & 2.0 &\\
 0.04 & 2.9 & 1.0 & 2.1 &`weak' at best\\
 0.01 & 8.0 & 2.1 & 2.6 &\\
  0.006 & 12 & 2.5 & 2.7&`moderate' at best\\
 0.003 & 21 & 3.0 & 3.0& \\
 0.001 & 53 & 4.0 & 3.3 &\\
 0.0003 & 150 & 5.0 & 3.6&`strong' at best\\
 $6\times 10^{-7}$ &
   43000 & 11 & 5.0&\end{tabular}
\end{center}
\caption{\label{translation} Translation table (using \eq{calib})
between p--values and the upper bounds on the odds  ($\bB$)
between the two models.  The `sigma' column is the corresponding
number of standard deviations away from the mean for a normal
distribution. In the `category' column are the descriptions for
the different categories of support reachable for the
corresponding p--value.
}
\end{table}
Note how the p--value of 0.05 (a 95\% confidence level result)
only corresponds to an odds ratio upper bound of $\bB=2.5$ and so
does not quite reach the ``weak'' support  threshold even for an
optimized prior. Also note that in order order for the ``strong''
support threshold to be reachable, $\sigma \ge 3.6$ is required.

 In general, for large sample size
and under mild regularity conditions, the p--value for the
addition of one or more new parameters can be estimated by finding
the maximum likelihood with the new parameters fixed
($\Lmax^{*}$), and when the new parameters are allowed to vary
($\Lmax$). Then the quantity \beq
\dchieff\equiv-2\ln(\Lmax^{*}/\Lmax) \eeq has a Chi squared
distribution with the number of degrees of freedom equal to the
number of new parameters \citep{wilks38}. It is important to note
that for this to be valid none of the new parameters can have
their fixed values on the boundary of the parameter space
 (see e.g.
\cite{Protassov:2002sz} for an astronomy-oriented example where
this condition does not hold). The p--value can then be estimated
by
 \beq \p =
\int_{y=\dchieff}^{\infty }
\chi^{2}_{\nu}(y) \, dy = 1-{\Gamma(\nu/2,\dchieff/2)\over\Gamma(\nu/2)}
   \label{pval}
\eeq where $\chi^{2}_\nu$ is the Chi squared distribution with
$\nu$ degrees of freedom, and $\nu$ is the number of new
parameters. Eq.~(\ref{pval}) is simply the asymptotic probability
of obtaining a $\Delta \chi^2$ as large or larger then what has
actually been observed, $\dchieff$, assuming the null hypothesis
is true. If the above procedure cannot be applied (for instance
because the new parameters lie at a boundary of the parameter
space), then the p--value can still be obtained by Monte Carlo
simulations.

A very different approach to estimating the Bayes factor without
having to specify a prior is the Bayesian Information Criteria
(BIC)
 \citep{schwarz78,magsor06,liddle07}. The BIC assumes a prior for the new parameters which
is equivalent to a single data point \citep{raftery95}. Therefore,
it will in general give lower values for $B$.
The BIC is complementary to the upper bound for $B$ presented here in that it provides a default weak rather than default strong prior.

\section{An illustrative Example}

Consider the case where under $M_0$, $x\sim\N(\mu_0,\sigma)$ for
fixed $\mu_{0}$ (the null hypothesis), while under the alternative
$M_1$, $x\sim\N(\mu,\sigma)$ and $N$ data samples are available
(with $\sigma$ known). If the prior on $\mu$ is taken to be
symmetric about $\mu=\mu_0$ and unimodal, then \citep{bersel87}
\beq \bB = { \phi(K+t) + \phi(K-t) \over 2 \phi(t) }
\label{example}
 \eeq where
$t\equiv \sqrt{N}|\bar{x}-\mu_{0}|/\sigma$, $\phi(y) \equiv
\e^{-y^2/2}$, and $K$ is found by solving \beq
K[\phi(K+t)+\phi(K-t)]=\int_{-(K+t)}^{K-t} \phi(y)\, \d y\, . \eeq
Alternatively, the p--value is given by
 \beq
 \label{computepval}
\p=1-\int_{y=-\tobs}^{\tobs}\phi(y)\, \d y\,.
 \eeq This can be
converted to a upper bound on the Bayes factor using \eq{calib}.
The results for the two methods are virtually identical and can be
read off Table~\ref{translation} where $t$ is the number of sigma.

\citet{selbayber01} present an interesting simulation study of
this model.
Consider the case described above, and let us generate data from a
random sequence of null hypothesis ($M_0$) and alternatives
($M_1$), with $\mu_0=0$, $\sigma=1$ and $\mu \sim \N(0,1)$.
Suppose that the proportion of nulls and alternatives is equal. We
then compute the p--value using Eq.~(\ref{computepval}) and we
select all the tests that give $\p  \in
[\alpha-\epsilon,\alpha+\epsilon]$, for a certain value of
$\alpha$ and $\epsilon\ll\alpha$.
Among such results, which rejected the null hypothesis at the
$1-\alpha$ level, we then determine the proportion that actually
came from the null, i.e. the percentage of wrongly rejected nulls.
We assume that either $M_1$ or $M_{0}$ is true. This allows us to
use \beq P(M_0|x)={1\over 1+B}\, . \label{justtwo} \eeq The
results are shown in Table~\ref{tab:nulls}. We notice that among
all the ``significant'' effects at the $2\sigma$ level  about 50\%
are wrong, and in general when there is only a single alternative
at least 29\% of the $2\sigma$ level results will be wrong.

\begin{table}
\begin{center}
\begin{tabular}{l|l l|l|}
p--value         &  sigma &  fraction of true nulls    &  lower
bound
                \\
\hline
0.05            &   1.96 & $0.51$ &  0.29\\
0.01            &   2.58 & $0.20$ &  0.11   \\
0.001           &   3.29 & $0.024 $ & 0.018        \\
\end{tabular}
\end{center}
\caption{\label{tab:nulls} Proportion of wrongly rejected nulls
among all results reporting a certain p--value (simulation
results). This  illustrates that the p--value is
 not equal to the fraction of wrongly rejected true nulls, which can
be considerably worse. This effect does not depend on the
assumption of Gaussianity nor on the sample size. The right most
column gives a lower bound on the fraction of true nulls derived
using \eqs{calib}{justtwo}.}
\end{table}

The root of this striking disagreement with a common
misinterpretation of the p--value (namely, that the p--value gives
the fraction of wrongly rejected nulls in the long run) is
twofold. While the p--value gives the probability of obtaining
data that are as extreme or more extreme than what has actually
been observed {\em assuming the null hypothesis is true}, one is
not allowed to interpret this as the probability of the null
hypothesis to be true, which is actually the quantity one is
interested in assessing. The latter step requires using Bayes
theorem and is therefore not defined for a frequentist. Also,
quantifying how rare the observed data are under the null is not
meaningful unless we can compare this number with their rareness
under an alternative hypothesis. Both these points are discussed
in greater detail in \cite{bersel87,selbayber01,berger02}.

\section{Scalar Spectral Index}
\label{sec:n}

Here we evaluate the upper bounds on the Bayes factor for the
scalar spectral index ($n$) using  WMAP combined with other data,
comparing a Harrison--Zeldovich model ($n=1$) to a model where $n$
can assume other values. For this problem, there are well
motivated priors.
If the primordial perturbations are from slow roll inflation, then 
 \beq n = 1+2\eta-6\epsilon \eeq where $\eta$ and $\epsilon$ are
 the slow roll parameters and need to be much less than one. For
most models $\epsilon\ll\eta$ and so a reasonable prior bound is
\beq 0.8 \lsim n\lsim 1.2 \, ,  \eeq
 which can be implemented by taking a Gaussian prior of the form
 \beq \label{nprior}
 p(n_s|M_1) = \N(\mu=1.0,\sigma= 0.2).
 \eeq
 However, if the inflation potential ($V$) is of the form
\beq V=V_{0}-{1\over 2}m^{2}\phi^{2} \eeq (where $\phi$ is the
inflaton, and $V_{0}$ and $m$ are constants) then inflation can
occur with $\eta\sim 1$ \citep{linde01,boulyt05} and so a larger
range of $n$ may be considered for the prior.

As there is such a broad range for the the prior on $n$, it is
useful to evaluate what is the upper bound on the odds for a
non--Harrison--Zeldovich spectrum, $n\neq 1$. In Table~\ref{n} we
list a number of different studies of the variation of the
spectral index for a range of data. Where the Bayes factor has
been worked out it can be seen that our estimate of the upper
bound is always more than the evaluated version. Also, for the
case with the greatest amount of data there is quite a large
discrepancy between the upper bound and the evaluated odds. This
makes sense as the same prior for $n$ was used (\eq{nprior}) but
now the data is more constraining and so the maximizing prior is
narrower. Using the most constraining data combination
(WMAPext+HST+SDSS) the upper limits on the odds against $n=1$ is
49:1. However, the odds against Harrison--ZelÕdovich could  be
weakened by various systematic effects in data analysis choices,
e.g. inclusion of gravitational lensing, beam
modelling, not including
Sunyaev-Zeldovich (SZ) marginalization,
 and point-source subtraction
\citep{peieas06,lewis06,parmuklid06,eriksen06,huferihan06,WMAP06}.
\begin{table}
\begin{center}
\begin{tabular}{lrrrrr}
data                           & $\dchieff$  & p--value& $\ln B$ &
$\ln \bB$ & $\overline{B}$ \\ \hline
WMAP                       &        6          &0.014    &  --     & 1.8 & 6\\
\citep{WMAP06}        &                   &               &         &       \\\hline
WMAPext+SDSS       &        8          &0.005    & 2.0    & 2.7 & 15\\
+2df+No SZ     &                   &              &          &        \\
  \citep{parmuklid06}    &                   &              &          &        \\ \hline
WMAPext+HST         &        8          &0.004    & 2.7    & 2.8 & 16\\
\citep{kuntropar06}    &                   &              &          &        \\ \hline
WMAPext+HST+SDSS&       11      &  0.001   & 2.9    & 3.9 & 49\\
 \citep{trotta05}          &                   &              &           &
\end{tabular}
\end{center}
\caption{ \label{n} The odds against a Harrison--ZelÕdovich
spectrum. The p--values where estimated from $\dchieff$ using
\eq{pval}. The upper bounds on the Bayes factor were estimated
using \eq{calib}.  Where $\ln B$ is available it was calculated
with the prior of \eq{nprior}. }
\end{table}

\section{Asymmetry in the CMB}
\label{sec:as} In the recent WMAP 3--yr release  the isotropy of
the CMB fluctuations was tested using a dipolar modulating
function \citep{WMAP06} \beq \Delta T(\n) = \Delta T_{\rm iso}(\n)
(1+A\n\cdot\D) \label{modulation1} \eeq where $\Delta T$ is the CMB
temperature fluctuations in direction $\n$, $\Delta T_{\rm iso}$
are the underlying isotropically distributed temperature
fluctuations, $A$ is the amplitude of the isotropy breaking, and
$\D$ is the direction of isotropy breaking. The isotropy of the
fluctuations can then be tested by evaluating whether $A=0$. The
problem with using the Bayes ratio in this case is that there is
no good underlying model which produces this type of isotropy
breaking. An attempt was made by \citet{dondutros07} to allow an
initial gradient in the inflaton field but they found that the
modulation dropped sharply with scale. However, the required
modulation should probably extend all the way to scales associated
with the harmonic $\ell=40$ \citep{hanbangor04}. Also,
\citet{inosil06} postulated that Poisson distributed voids may be
responsible for the asymmetry. But, a generating mechanism for the
voids and a detailed likelihood analysis are presently lacking.

Therefore at present there is not a concrete enough theory to
place meaningful prior limits on $A$. However, we can still work
out the upper limit on the Bayes factor. The p--values can be
evaluated from \eq{pval}. Although $A=0$ is on the boundary of the
parameter space, the problem can be reparameterized in Cartesian
coordinates where $A=w_{x}^{2}+w_{y}^{2}+w_{z}^{2}$ and $w_{i}$ is
a linear modulation weight for spatial dimension $i$. Then the
$w_{i}=0$ point, for all $i$, will not be on the edge of the
parameter space and so \eq{pval} can be used.

The results are shown in Table~4. Simulations had been done for
the last row's p--value \citep{eriksen07} and were in excellent
agreement with the result from \eq{pval}.
 \citet{eriksen07} did compute the Bayes factor, taking as the prior  $0\le A\le0.3$
but did not give a justification for that prior except that it contained
all the non-negligible likelihood. This is unproblematic for
parameter estimation, but is ambiguous for working out the Bayes
factor. For example if the prior range for  $A$ was extended to be
$0\le A \le 0.6$ then the Bayes factor would decrease by 2 but the
parameter estimates would be unaffected.
\begin{table}
\begin{center}
\begin{tabular}{lrrrrr}
data & $\dchieff$  & p--value  & $\ln B$ & $\ln \bB$ &
$\overline{B}$\\ \hline
WMAP ($7^{\circ}$)                      &        3         &0.4         & --          & -- & -- \\
 \citep{WMAP06}                          &                   &              &             &     \\ \hline
WMAP ($7^{\circ}$)+$C_{\rm marg}$  &       9          &0.03        & --         & 1.3  & 4\\
\citep{gordon07}                              &                     &                 &            &       \\ \hline
WMAP ($3.6^{\circ}$)+$C_{\rm marg}$&       11         &0.01       & 1.8       &  2.16 & 9\\
\citep{eriksen07}                               &                      &              &               &
\end{tabular}
\end{center}
\caption{ \label{modulation} The odds for dipolar modulation,
$A\neq0$. The resolution of the data used is also indicated. The
$C_{\rm marg}$ refers to marginalisation over a non-modulated
monopole and dipole. 
 $\ln \bB$ was
evaluated using \eq{calib}. }
\end{table}

\section{Conclusions}

Bayesian model selection provides a powerful way of evaluating
whether new parameters are needed in a model. There are however
cases where the prior for the new parameter can be uncertain, or
physically difficult to motivate. Here we have looked at priors
which maximize the Bayes factor for the new parameters. This puts
the reduced model under the most strain possible and so tells the
user what the best case scenario is for the new parameters. We
have pointed out a common misinterpretation of the meaning of
p--values, which often results in an overestimation of the true
significance of rejection tests for null hypotheses.

Using Bayesian calibrated p--values we have evaluated upper bounds
on the Bayes factor for the spectral index. We have found that the
best the current data can do is provide moderate support
(odds~$\le49:1$) for $n\not =1$. We also looked at the maximum
Bayes factor for a modulation in the WMAP CMB temperature data. We
found that the current data can at best provide weak support
(odds~$\le 9:1$) for a departure from isotropy.

The comparison between p--values and Bayes factors suggests a
threshold of $\p= 3 \times 10^{{-4}}$ or $\sigma = 3.6$ is needed
if the odds of 150:1 (``strong'' support at best) are to be
obtained. It is difficult to detect systematics which are smaller
than the statistical noise and so systematic effects in the data
analysis typically lead to a shift of order a sigma. It follows
that the ``particle physics discovery threshold'' of
 $5\sigma$ may be required in order to obtain odds of at best 150:1.

\section*{Acknowledgments}

We are grateful to Kate Land for useful conversations and to Uros
Seljak for interesting comments. CG is supported by Beecroft
Institute for Particle Astrophysics and Cosmology. RT is supported
by the Royal Astronomical Society through the Sir Norman Lockyer
Fellowship, and by St Anne's College, Oxford.

\bibliographystyle{mn2e_eprint}

\bibliography{bib}

\begin{thebibliography}{}

\bibitem[\protect\citeauthoryear{Berger}{Berger}{2003}]{berger02}
Berger J.,  2003, Statistical Science, 18, 1

\bibitem[\protect\citeauthoryear{Berger \& Sellke}{Berger \&
  Sellke}{1987}]{bersel87}
Berger J.~O.,  Sellke T.,  1987, The American Statistician, 55, 62

\bibitem[\protect\citeauthoryear{{Boubekeur} \& {Lyth}}{{Boubekeur} \&
  {Lyth}}{2005}]{boulyt05}
{Boubekeur} L.,  {Lyth} D.~H.,  2005, Journal of Cosmology and Astro-Particle
  Physics, 7, 10, \adsurl{http://adsabs.harvard.edu/abs/2005JCAP...07..010B},
  \eprint{arXiv:hep-ph/0502047}

\bibitem[\protect\citeauthoryear{Donoghue, Dutta \& Ross}{Donoghue
  et~al.}{2007}]{dondutros07}
Donoghue J.~F.,  Dutta K.,    Ross A.,  2007, ArXiv Astrophysics e-prints,
  astro-ph/0703455, \eprint{astro-ph/0703455}

\bibitem[\protect\citeauthoryear{{Eriksen}, {Banday}, {Gorski}, {Hansen} \&
  {Lilje}}{{Eriksen} et~al.}{2007}]{eriksen07}
{Eriksen} H.~K.,  {Banday} A.~J.,  {Gorski} K.~M.,  {Hansen} F.~K.,    {Lilje}
  P.~B.,  2007, ApJ, 660, L81,
  \adsurl{http://adsabs.harvard.edu/abs/2007astro.ph..1089E},
  \eprint{astro-ph/0701089}

\bibitem[\protect\citeauthoryear{{Eriksen} et~al.,}{{Eriksen}
  et~al.}{2007}]{eriksen06}
{Eriksen} H.~K.,  et~al., 2007, ApJ, 656, 641,
  \adsurl{http://adsabs.harvard.edu/abs/2007ApJ...656..641E},
  \eprint{arXiv:astro-ph/0606088}

\bibitem[\protect\citeauthoryear{Feroz \& Hobson}{Feroz \&
  Hobson}{2007}]{Feroz:2007kg}
Feroz F.,  Hobson M.~P.,  2007, ArXiv Astrophysics e-prints, arXiv:0704.3704,
  \eprint{arXiv:0704.3704 [astro-ph]}

\bibitem[\protect\citeauthoryear{{Gordon}}{{Gordon}}{2007}]{gordon07}
{Gordon} C.,  2007, ApJ, 656, 636,
  \adsurl{http://adsabs.harvard.edu/abs/2007ApJ...656..636G},
  \eprint{arXiv:astro-ph/0607423}

\bibitem[\protect\citeauthoryear{Hansen, Banday \& Gorski}{Hansen
  et~al.}{2004}]{hanbangor04}
Hansen F.~K.,  Banday A.~J.,    Gorski K.~M.,  2004, MNRAS, 354, 641,
  \eprint{astro-ph/0404206}

\bibitem[\protect\citeauthoryear{{Huffenberger}, {Eriksen} \&
  {Hansen}}{{Huffenberger} et~al.}{2006}]{huferihan06}
{Huffenberger} K.~M.,  {Eriksen} H.~K.,    {Hansen} F.~K.,  2006, ApJ, 651,
  L81, \adsurl{http://adsabs.harvard.edu/abs/2006ApJ...651L..81H},
  \eprint{arXiv:astro-ph/0606538}

\bibitem[\protect\citeauthoryear{{Inoue} \& {Silk}}{{Inoue} \&
  {Silk}}{2006}]{inosil06}
{Inoue} K.~T.,  {Silk} J.,  2006, ApJ, 648, 23,
  \adsurl{http://adsabs.harvard.edu/abs/2006ApJ...648...23I},
  \eprint{arXiv:astro-ph/0602478}

\bibitem[\protect\citeauthoryear{{Kunz}, {Trotta} \& {Parkinson}}{{Kunz}
  et~al.}{2006}]{kuntropar06}
{Kunz} M.,  {Trotta} R.,    {Parkinson} D.~R.,  2006, PRD, 74, 023503,
  \adsurl{http://adsabs.harvard.edu/abs/2006PhRvD..74b3503K},
  \eprint{arXiv:astro-ph/0602378}

\bibitem[\protect\citeauthoryear{{Lewis}}{{Lewis}}{2006}]{lewis06}
{Lewis} A.,  2006, ArXiv Astrophysics e-prints, astro-ph/0603753,
  \adsurl{http://adsabs.harvard.edu/abs/2006astro.ph..3753L},
  \eprint{astro-ph/0603753}

\bibitem[\protect\citeauthoryear{{Liddle}}{{Liddle}}{2007}]{liddle07}
{Liddle} A.~R.,  2007, Mon. Not. Roy. Astron. Soc. Lett., 377, L74,
  \adsurl{http://adsabs.harvard.edu/abs/2007astro.ph..1113L},
  \eprint{astro-ph/0701113}

\bibitem[\protect\citeauthoryear{{Linde}}{{Linde}}{2001}]{linde01}
{Linde} A.,  2001, Journal of High Energy Physics, 11, 52,
  \adsurl{http://adsabs.harvard.edu/abs/2001JHEP...11..052L},
  \eprint{arXiv:hep-th/0110195}

\bibitem[\protect\citeauthoryear{{Magueijo} \& {Sorkin}}{{Magueijo} \&
  {Sorkin}}{2007}]{magsor06}
{Magueijo} J.,  {Sorkin} R.~D.,  2007, Mon. Not. Roy. Astron. Soc. Lett., 377,
  L39, \adsurl{http://adsabs.harvard.edu/abs/2006astro.ph..4410M},
  \eprint{astro-ph/0604410}

\bibitem[\protect\citeauthoryear{{Mukherjee}, {Parkinson} \&
  {Liddle}}{{Mukherjee} et~al.}{2006}]{mukparlid06}
{Mukherjee} P.,  {Parkinson} D.,    {Liddle} A.~R.,  2006, ApJ, 638, L51,
  \adsurl{http://adsabs.harvard.edu/abs/2006ApJ...638L..51M},
  \eprint{arXiv:astro-ph/0508461}

\bibitem[\protect\citeauthoryear{Parkinson, Mukherjee \& Liddle}{Parkinson
  et~al.}{2006}]{parmuklid06}
Parkinson D.,  Mukherjee P.,    Liddle A.~R.,  2006, Phys. Rev., D73, 123523,
  \eprint{astro-ph/0605003}

\bibitem[\protect\citeauthoryear{{Peiris} \& {Easther}}{{Peiris} \&
  {Easther}}{2006}]{peieas06}
{Peiris} H.~V.,  {Easther} R.,  2006, Journal of Cosmology and Astro-Particle
  Physics, 10, 17, \adsurl{http://adsabs.harvard.edu/abs/2006JCAP...10..017P},
  \eprint{arXiv:astro-ph/0609003}

\bibitem[\protect\citeauthoryear{Protassov, van Dyk, Connors, Kashyap \&
  Siemiginowska}{Protassov et~al.}{2002}]{Protassov:2002sz}
Protassov R.,  van Dyk D.~A.,  Connors A.,  Kashyap V.~L.,    Siemiginowska A.,
   2002, ArXiv Astrophysics e-prints, astro-ph/0201547,
  \eprint{astro-ph/0201547}

\bibitem[\protect\citeauthoryear{Raftery}{Raftery}{1995}]{raftery95}
Raftery A.,  1995, Sociological Methodology, 25, 11

\bibitem[\protect\citeauthoryear{Schwarz}{Schwarz}{1978}]{schwarz78}
Schwarz G.,  1978, Annals of Statistics, 6, 461

\bibitem[\protect\citeauthoryear{Sellke, J. \& Berger}{Sellke
  et~al.}{2001}]{selbayber01}
Sellke T.,  J. B.~M.,    Berger J.~O.,  2001, Journal of the American
  Statistical Association, 82, 112

\bibitem[\protect\citeauthoryear{Spergel et~al.,}{Spergel
  et~al.}{2007}]{WMAP06}
Spergel D.~N.,  et~al., 2007, ApJS, 170, 377, \eprint{astro-ph/0603449}

\bibitem[\protect\citeauthoryear{{Trotta}}{{Trotta}}{2007}]{trotta05}
{Trotta} R.,  2007, Mon. Not. R. Astron. Soc., 378, 72,
  \adsurl{http://adsabs.harvard.edu/abs/2005astro.ph..4022T},
  \eprint{astro-ph/0504022}

\bibitem[\protect\citeauthoryear{Trotta}{Trotta}{2007a}]{Trotta:2007hy}
Trotta R.,  2007a, MNRAS, 378, 819, \eprint{astro-ph/0703063}

\bibitem[\protect\citeauthoryear{Trotta}{Trotta}{2007b}]{Trotta:2006ww}
Trotta R.,  2007b, Mon. Not. Roy. Astron. Soc. Lett., 375, L26,
  \eprint{astro-ph/0608116}

\bibitem[\protect\citeauthoryear{Wilks}{Wilks}{1938}]{wilks38}
Wilks S.~S.,  1938, The Annals of Mathematical Statistics, 9, 60

\end{thebibliography}

\end{document}